\documentclass[twocolumn]{aastex631}

\renewcommand{\bm}[1]{{\mbox{{\boldmath$#1$}}}}	

\newcommand{\M}[1]{\mathbf{#1}}

\newcommand{\grad}{\bm{\nabla}}

\usepackage{amsmath}

\begin{document}

\title{Evolution of a Long-Lived Deep-Seated Main-Sequence Magnetic Field During White Dwarf Cooling}

\author[0000-0003-4636-7898]{Matias Castro-Tapia}
\altaffiliation{Co-first authorship. These authors contributed equally to this work.}
\affiliation{Department of Physics and Trottier Space Institute, McGill University, Montreal, QC H3A 2T8, Canada}
\email{matias.castrotapia@mail.mcgill.ca}

\author[0000-0002-3524-190X]{Maria Camisassa}
\altaffiliation{Co-first authorship. These authors contributed equally to this work.}
\affiliation{Departament de Física, Universitat Politècnica de Catalunya, c/Esteve Terrades 5, 08860 Castelldefels, Spain}
\email{maria.camisassa@upc.edu}

\author[0009-0009-8666-9218]{Shu Zhang}
\affiliation{Department of Physics and Trottier Space Institute, McGill University, Montreal, QC H3A 2T8, Canada}



\begin{abstract}

We study the evolution of white dwarf (WD) magnetic fields that originate from core-convective dynamos during the main-sequence. Using stellar evolution and WD cooling models combined with magnetic field diffusion calculations, we demonstrate that a surviving field from the main-sequence can account for various features observed in magnetic WDs. In particular, the earlier emergence of stronger magnetic fields in more massive WDs, compared to older, less massive, and less magnetic ones, can be explained by this framework. This is because the magnetic boundary at the onset of WD cooling lies deeper in less massive WDs, resulting in a slower and weaker evolution of the surface magnetic field due to increasing electrical conductivity over time. We further show that many of the magnetic field strengths observed across different WD samples can be reproduced if the deep-seated field generated during the main sequence is comparable to predictions from magnetohydrodynamic simulations of core-convective dynamos, or if equipartition provides a valid scaling for the main-sequence dynamo. Additionally, our predictions for surface magnetic fields vary by a factor of 2 to 4 when higher-order modes of poloidal magnetic field expansion and turbulent diffusion driven by crystallization-induced convection are included. These effects should therefore be considered when investigating the origin of magnetic fields in individual WDs.
\end{abstract}


\keywords{ Stellar evolution(1599) --- Stellar magnetic fields(1610) ---- White dwarf stars(1799) ---
Stellar interiors(1606)}


\section{Introduction} \label{sec:intro}

White dwarf (WD) stars represent the most common endpoint of stellar evolution, as all main sequence stars with masses below 9-12M$_\odot$ will eventually become WDs. Consequently, these old and numerous objects provide valuable information on the stellar evolution theory, the final states of planetary systems, the structure and evolution of our Galaxy, and the properties of stellar matter under extreme conditions ---  see  for  instance, the  reviews  of
\cite{2008PASP..120.1043F},       \cite{2008ARA&A..46..157W},      
\cite{2010A&ARv..18..471A}, \cite{2019A&ARv..27....7C}, \cite{2022FrASS...9....6I},  \cite{2024NewAR..9901705T}, and \cite{2025AN....34640118C}.
In fact, WDs preserve valuable information about the evolutionary history of their progenitor stars and can be used to constrain nuclear reaction rates \citep{2019A&A...630A.100D}, the initial-to-final mass relation \citep{2008MNRAS.387.1693C,2018ApJ...866...21C}, and the occurrence of third dredge-up episodes in the Asymptotic Giant Branch (AGB) phase \citep{2015A&A...576A...9A}, among other key processes.

Magnetic fields have been consistently detected on the surface of WDs for the past 55 years \citep{1970ApJ...161L..77K,1970ApJ...160L.147A}, and today more than 600 magnetized WDs are known. These fields have been observed across a diverse range of WD types, including both single and binary systems, spanning various spectral classes, masses, and evolutionary stages \citep[see][for reviews]{2015SSRv..191..111F,2020AdSpR..66.1025F}. The strength of these magnetic fields varies widely, from $10^3$ to $10^9$  G.
Recent studies of the volume-limited sample of magnetic WDs within 20 parsecs of the Sun have contributed significantly to our understanding of these objects \citep{2007ApJ...654..499K,2021MNRAS.507.5902B}. These studies have systematically examined each of the WDs in this region searching for magnetic fields, thus eliminating the observational biases inherent in previous magnitude-limited surveys. Their findings reveal that magnetic WDs tend to be, on average, more massive than their non-magnetic counterparts. Moreover, they discovered a strong correlation between magnetism and core crystallization, with a higher prevalence of magnetic fields in WDs that have undergone or are currently undergoing the crystallization process.
Further expanding on this work, \cite{2022ApJ...935L..12B} extended the sample to 40 parsecs, but only considering those WDs younger than 0.6 gigayears (Gyr). Their analysis reaffirmed previous conclusions, confirming that the occurrence of magnetism in young (pre-crystallization) WDs is approximately 10\%, whereas it increases to around 30\% in older, crystallized WDs.

Despite the extensive observational data available of magnetic WDs, the origin and evolution of these magnetic fields remain uncertain. Theories involving binary and single evolution have been proposed, but a definitive explanation has not yet been established. One possibility is that the field is generated as a result of close binary interactions, either during a merger episode \citep{2012ApJ...749...25G}, or during a post-common-envelope phase \citep{2008MNRAS.387..897T}. Another possibility, the so-called fossil field hypothesis, suggests that WDs inherited a magnetic field from the stellar formation phase. This hypothesis is also linked with the magnetic fields observed in the surface of peculiar Ap and Bp stars, which can later be inherited by their WD descendants, being their magnetic field strengths consistent when assuming a simple model of magnetic flux conservation.
More recently, \citet{2017ApJ...836L..28I} proposed that the compositionally driven mixing induced during WD crystallization can generate a dynamo magnetic field. Although this hypothesis a priori seems promising, both the predicted strength and the time of emergence of this crystallization-driven magnetic field struggle to match consistently the observations of magnetic WDs \citep{2024MNRAS.528.3153B, 2024MNRAS.533L..13B, Fuentes2024, 2024ApJ...961..197M, Castro-Tapia2024a, Castro-Tapia2024b}.

A recent alternative explanation for the WD magnetic fields has been explored in \cite{2024A&A...691L..21C}, which proposes that magnetic fields arise as a result of the surface emergence of a deep-seated convective dynamo action. Main-sequence stars more massive than $\sim 1.1 M_\odot$ have convective cores that can sustain dynamo magnetic fields. This idea is supported by magnetohydrodynamic (MHD) simulations of convective cores of main sequence stars \citep[e.g.][]{2016ApJ...829...92A,2024A&A...691A.326H} and by the astroseismologic detection of magnetic fields buried in the deep interior of red giant stars \citep[e.g.][]{2016Natur.529..364S,2023A&A...670L..16D}. Specifically, \cite{2024A&A...691L..21C} studied the evolution of a main-sequence dynamo magnetic field through the subsequent evolutionary phases to the white dwarf phase. These authors found that, if this field is generated and can survive trapped in the stellar interior throughout the evolution, it would emerge to the surface during the WD stage. Moreover, the time of surface emergence of the main-sequence dynamo can explain most of the magnetic WDs with masses $\gtrsim 0.65M_\odot$. However, this work relies on a basic estimation of the time of emergence through magnetic diffusion, and this basic picture cannot predict the magnetic field strength at the WD surface. 

This paper is precisely aimed at studying in detail the diffusion of a deep-seated main-sequence dynamo magnetic field during the white dwarf phase. To such end, we solve the induction equation for the WD interior assuming an initial axisymetric poloidal field buried below the surface as predicted in \cite{2024A&A...691L..21C}. In this way, assuming an initial magnetic field strength configuration for the interior, we can obtain the time evolution of the magnetic field on the WD surface as a result of diffusion. Or, conversely, knowing the magnetic field strength at the WD surface, we can estimate the initial field strength generated during the main sequence phase. We obtained that the magnetic field strengths in the WD phase are compatible with the predictions for the main sequence dynamo based on MHD models. We have also estimated the magnetic field strength of the main sequence dynamo based on kinetic and magnetic energy equipartition, and we found that these are also consistent with the fields observed in the WD phase.

\section{Methods}

\subsection{Main sequence dynamo and magnetic boundaries}

We assumed that during the main sequence, stars that develop a convective core can sustain a magnetic dynamo. This assumption is based on MHD simulations of convective cores of main-sequence stars. For instance, \citet{2005ApJ...629..461B} performed three-dimensional
nonlinear simulations of 2 $M_\odot$ A-type stars with rotation rates of 1 and 4 times the solar mean angular velocity, showing that vigorous convection
can amplify the initial seed fields and sustain them against ohmic decay. Dynamo action can yield magnetic fields that possess energy densities comparable to those of the convective motions. More recently, \cite{2024A&A...691A.326H}  performed simulations of a 2.2 $M_\odot$ A-type star with similar rotation rates, finding similar results. The magnetic field strengths obtained in these simulations are of the order of 60kG. For more massive stars, \cite{2016ApJ...829...92A}, simulated a  10 $M_\odot$ B-type star with different rotating rates, finding that the vigorous dynamo action can build magnetic fields strengths exceeding a $10^6$ G, and that magnetic energy in the faster rotators can reach super-equipartition. However, it is not clear whether a  10 $M_\odot$ star will evolve to a WD or it will explode as a core-collapse supernova. 

\begin{figure}[ht!]
\includegraphics[trim=0 0 0 0,clip,width=0.99\columnwidth]{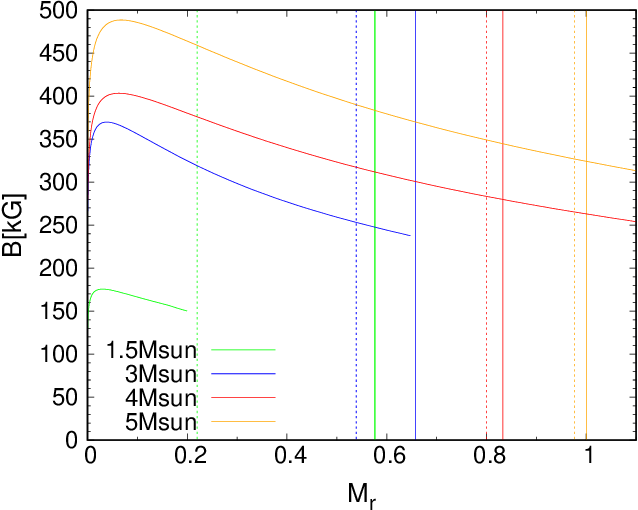}
\caption{Magnetic field strength of the main sequence dynamo assuming equipartition for main sequence stars of 1.5, 3, 4, and 5 M$_\odot$. The solid vertical lines indicate the final white dwarf masses, assuming the initial-to-final-mass relation of \cite{2008MNRAS.387.1693C}. The dashed vertical lines indicate the location of the magnetic boundaries below which we expect the magnetic field to be buried at the beginning of the white dwarf phase, assuming the cubic fit prescription from \cite{2024A&A...691L..21C}. Note that in the x-axis $\mathrm{M_{r}}=m/M_{\odot}$, where $m$ is the total mass contained at the location from the center in solar masses.
\label{fig:equipartition}}
\end{figure}

We have estimated the magnetic field strengths for the convective cores of main-sequence stars by assuming equipartition between the magnetic energy and the kinetic energy of the convective flows. To such end, we have simulated main sequence stars with 1.5, 3, 4, and 5 $M_\odot$ using the {\tt LPCODE} stellar evolution code, developed by the La
Plata group \citep{2005A&A...435..631A,2015A&A...576A...9A,2016A&A...588A..25M}. We have estimated the field strength by $B=\sqrt{4\pi\rho v_{\rm c}^2}$, where $\rho$ is the density, and $v_{\rm c}$ is the characteristic convective velocity, which we have assumed as $v_{\rm c}\sim (F/\rho)^{1/3}$, and $F$ is the heat flux. 
The resulting magnetic fields are shown in Figure \ref{fig:equipartition}, where the vertical solid lines indicate the final WD mass according to the initial-to-final-mass relation of \cite{2008MNRAS.387.1693C} and the dashed vertical lines show the location of the magnetic boundaries below which we expect the magnetic field to be buried at the beginning of the white dwarf phase. The magnetic boundary is taken from the cubic fit prescription of \cite{2024A&A...691L..21C}. This boundary is set at the maximum extension of the convective core during the main sequence (in those stars where the subsequent outer convective zones during the first and second dredge-up do not reach this boundary) or by the boundary of the maximum extension of the outer convective zone during the first and second dredge-up episodes (in those stars where the first or second dredge-up reach the regions that have previously been part of the convective core) \citep[see][for details]{2024A&A...691L..21C}.

\subsection{White dwarf models}\label{sec:2.2}

To evaluate the evolution in the WD phase of the magnetic field that survived from the main-sequence, we employed the WD models with carbon-oxygen (C/O) cores and masses 0.6, 0.7, 0.8, 0.9, and 1.0 $M_\odot$ of \citet{Castro-Tapia2024b}, which are based on the work of \citet{Bauer2023}. The cooling of these WD models is computed using MESA version r24.08.1 \citep{Paxton2011, Paxton2013, Paxton2015, Paxton2018, Paxton2019, Jermyn2023} including C/O phase separation in the core as presented by \citet{Bauer2023}. For simplicity, we only included C, O, helium (He), and hydrogen (H) (with nuclear reactions turned
off) in the nuclear network.

To later include the effect of turbulent diffusion driven by compositional convection during crystallization, we calculated the location of the solid core $R_{\mathrm{core}}$ and the outermost location of the convection zone $R_{\mathrm{out}}$ at each time in each evolution profile. Specifically, we computed $R_{\mathrm{core}}$ following the work of \citet{Castro-Tapia2024a}, who used the phase parameter $\phi>0.9$ given by the Skye equation of state in MESA \citep{Jermyn2021}. To obtain  $R_{\mathrm{out}}$, we took the extension of the flat composition profile outside the solid core; such a profile is given by the mixing of the phase separation routine following the Ledoux criterion as shown in Figure 1 of \citet{Fuentes2024}.

\subsection{Magnetic Diffusion Calculation}\label{sec:2.3}
Since we have assumed that the magnetic field inside the WD originated in a previous stellar evolution phase, we use the induction equation, considering only the diffusion part, to calculate the field evolution as follows
\begin{equation}\label{eq_ind}
    \frac{\partial\bm{B}}{\partial{t}}=-\grad\times(\eta\grad\times\bm{B}),
\end{equation}
where $\eta$ is the magnetic diffusivity. Using the WD models described in Section \ref{sec:2.2} and the Electron Transport Coefficients of Magnetized Stellar Plasmas code of \citet{Potekhin1999, Potekhin2015}, we computed the electrical conductivities $\sigma$ along the star radial coordinate for different cooling times since the formation of the WD. Thus, in equation \eqref{eq_ind} we include the Ohmic diffusivity as $\eta_{\mathrm{ohm}}=c^{2}/4\pi\sigma$, where $c$ is the speed of light. As presented in \citet{Castro-Tapia2024b}, we used a free parameter $\mathrm{f_{Rm}}$ proportional to the magnetic Reynolds number to calculate the effect of turbulent diffusion as $\eta_{\mathrm{turb}}=\mathrm{f_{Rm}}\eta_{\mathrm{ohm}}$, and then the full conductivity including both contributions is $\eta=\eta_{\mathrm{ohm}}+\eta_{\mathrm{turb}}$.

Assuming an axisymmetric poloidal field and taking a separable magnetic vector potential $\bm{A}=A_{\phi}(r,\theta,t)\hat{e}_{\phi}=\sum_{l}[R_{l}(r,t)/r]P^{1}_{l}(\cos{\theta})\hat{e}_{\phi}$, where $R_{l}$ is the radial component and $P^{1}_{l}$ is the associated Legendre polynomial of order 1 (as done in \citealt{Castro-Tapia2024b}, motivated by previous work of \citealt{Cumming2002} and references therein), we use $\bm{B}=\grad\times\bm{A}$ to write
\begin{equation}\label{eq_Rl}
    \frac{\partial R_{l}}{\partial t}=\eta(r,t)\left[\frac{\partial^{2} R_{l}}{\partial r^{2}}-\frac{l(l+1)R_{l}}{r^{2}}\right].
\end{equation}
Thus, to obtain the diffusion of the magnetic field for our WD models with surviving fields from the core-convective dynamo, we solve numerically equation \eqref{eq_Rl} using the Crank–Nicolson method, taking as initial condition that the field is filling the WD interior up to the magnetic boundary computed from the fit of \citet{2024A&A...691L..21C}.
\section{Magnetic field evolution}
\subsection{Dipole and quadrupole components}\label{sec:3.1}
The magnetic fields observed in WDs are usually consistent with a dipole geometry with $l=1$ in the analytical expansion obtained in the previous section. However, more complex geometries are sometimes needed. For example, \citet{Hardy2023} and \citet{Moss2025} showed that an off-centered is a better fit for spectroctopic data of many MWDs. This result suggests that higher orders of $l$ should be considered when analyzing the magnetic field evolution. To address this problem, we considered the magnetic field diffusion for a dipole ($l=1$) only and for a dipole plus a quadrupole ($l=1,2$).

For the dipole only (d), we get that the components of the magnetic field are
\begin{equation}\label{eq_B_vec}
    \bm{B_{\mathrm{d}}}(r,\theta,t)=\frac{2R_{1}(r,t)}{r^{2}}\cos{\theta}\hat{e}_{r}+\frac{1}{r}\frac{\partial R_{1}(r,t)}{\partial{r}}\sin{\theta}\hat{e}_{\theta},
\end{equation}
while for the dipole plus quadrupole (d+q) we get 
\begin{align}\label{eq_B_vec_dq}
\nonumber \bm{B_{\mathrm{d+q}}}(r,\theta,t)=
\frac{1}{r^{2}}\left[2R_{1}(r,t)\cos{\theta}+3R_{2}(r,t)(3\cos^{2}{\theta}-1)\right]\hat{e}_{r}\\
+\frac{\sin{\theta}}{r}\left[\frac{\partial R_{1}(r,t)}{\partial{r}}+3\frac{\partial R_{2}(r,t)}{\partial{r}}\cos{\theta}\right]\hat{e}_{\theta}.
\end{align}
We use the boundary conditions $\partial R_{l}/\partial r=(l+1)R_{l}/r$ for the center of the star $r\rightarrow{0}$ and  $\partial R_{l}/ \partial r=-lR_{l}/r$ for $r\rightarrow{R_\mathrm{WD}}$, which enforces that the current fulfill $\bm{J}\propto\grad\times \bm{B}=0$ outside the star \citep{Cumming2002}.

\begin{figure*}[ht!]
\includegraphics[width=0.99\textwidth]{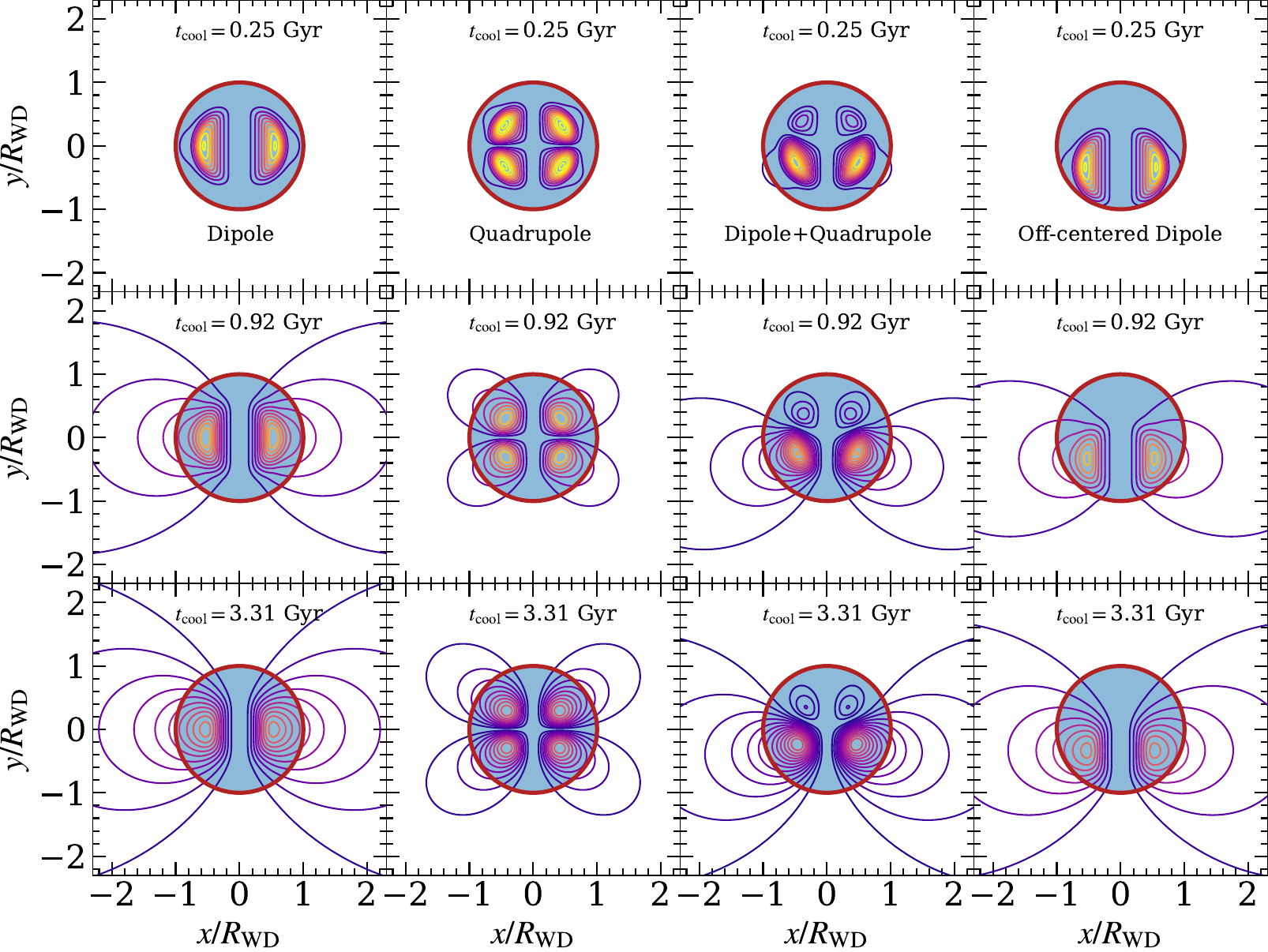}
\caption{Two-dimensional visualization of the evolution of the magnetic field lines for our 0.7 M$_\odot$ model, assuming an axisymmetric dipole field (left panels), a quadrupole field (middle-left panels), a dipole plus quadrupole field (middle-right panels), and an off-centered dipole field (right panels). The lines in each panel are the magnetic field lines with equal magnetic flux, and the color scaling indicates increasing magnetic field strength for brighter colors.
\label{fig:2d}}
\end{figure*}

For both cases, we set the initial magnitude of the root mean squared (rms) magnetic field over the solid angle $\Omega$ to be constant ($B_{0}$) inside the magnetic boundary. Thus,
\begin{align}
\nonumber B_{\mathrm{rms}}(r,t_{0})=\sqrt{\frac{1}{4\pi}\int_{4\pi}|\bm{B}|^{2}(r,\theta,t_{0})d
     \Omega}\\
     =
    \begin{cases} 
      B_{0} & 0\leq r \leq R_{\mathrm{MB}}\\
     0 & r > R_{\mathrm{MB}}
   \end{cases},
\end{align}
and for $\bm{B_\mathrm{d+q}}(r,\theta,t_{0})$ we also considered the initial condition $R_{1}(r,t_{0})=R_{2}(r,t_{0})$ for simplicity.

In Figure \ref{fig:2d}, we show the two-dimensional projections of the magnetic field lines evolution with equal magnetic flux for four different cases of a 0.7$M_{\odot}$ WD. In the left set of panels, we show the dipole (d) component only, in the second set the quadrupole (q) only, the third panels show the result of d+q, and the rightmost panels show a dipole off-centered in the vertical ($y$) direction by $-0.333 R_{\mathrm{WD}}$. The d+q and the off-centered dipole structure evolution are naturally very similar. This is expected, as the results shown by \citet{Hardy2023} and \citet{Moss2025}, where the spectra of many magnetic (M)WDs fit an off-centered dipole at $\sim0.3 R_{\mathrm{WD}}$, can be expanded by considering a d+q or higher order multipolar magnetic fields.

\begin{figure}[ht!]
\includegraphics[trim=0 0 0 0,clip,width=0.99\columnwidth]{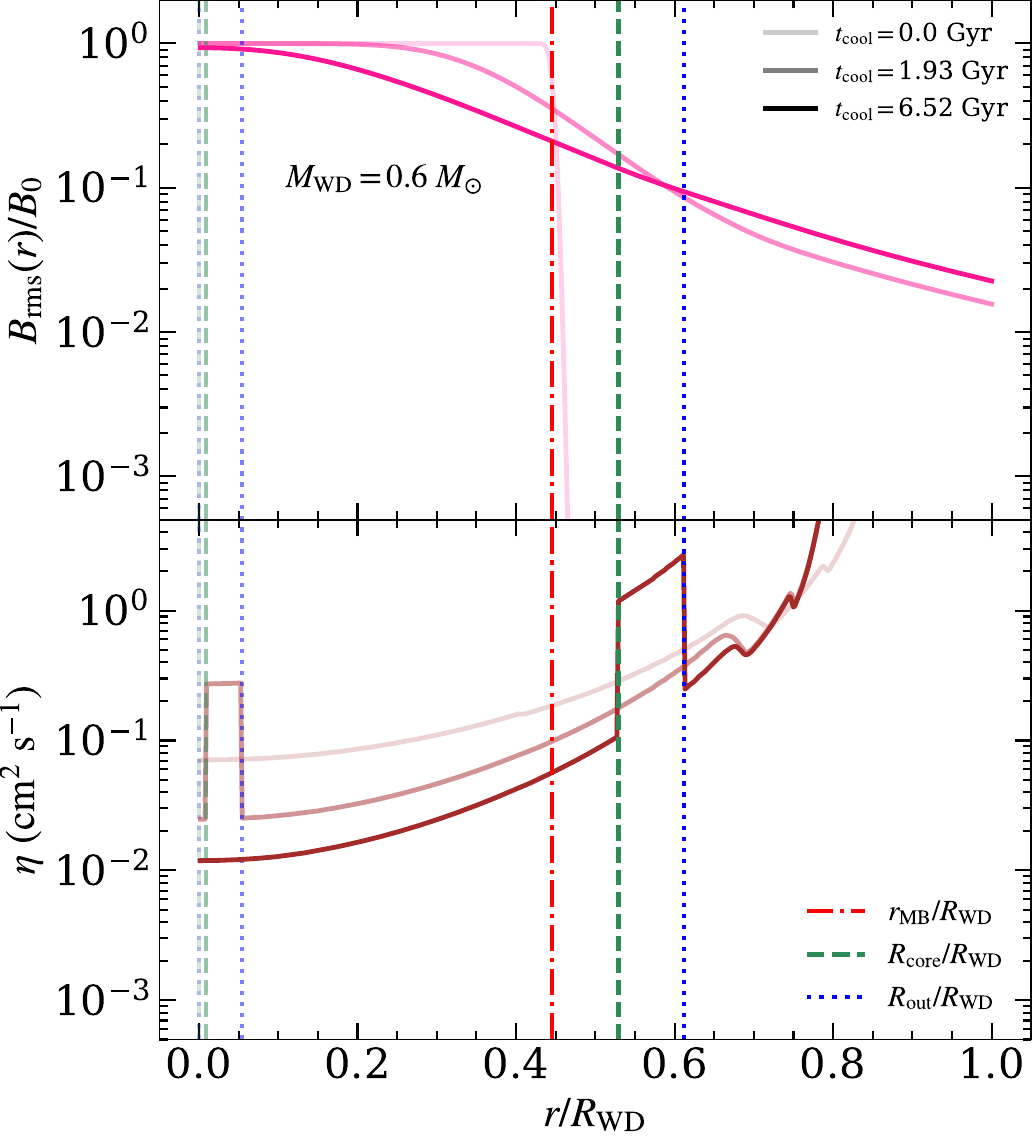}
\caption{Evolution of the magnetic field strength (top panel) and the magnetic diffusivity (bottom) as a function of the normalized white dwarf radius for our 0.6 M$_\odot$ model, assuming an initial axisymmetric dipole field. \textbf{In this case, we included turbulent diffusivity inside the compositionally-driven convection zone with $\mathrm{f_{RM}}=10$.}
\label{fig:evolution}}
\end{figure}

\begin{figure*}[ht!]
\includegraphics[width=0.99\textwidth]{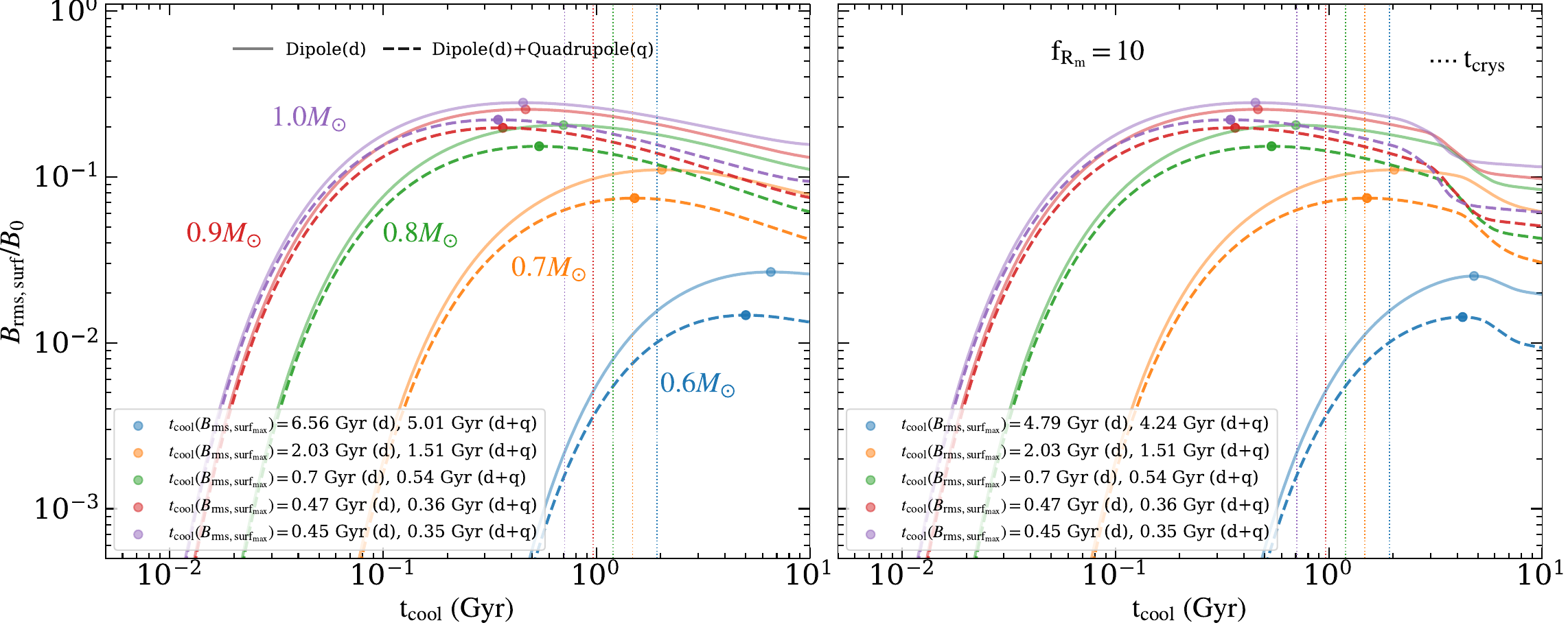}
\caption{Left panel: Surface magnetic field as a function of the cooling age for our dipole (solid lines) and dipole plus quadrupole (dashed lines) initial configurations. Right panel: Same as left panel, but simulating turbulent diffusion in the crystallization induced by Rayleigh-Taylor unstable regions (see text for details). Maximum values are indicated using filled circles, and the dotted vertical lines on the right panel indicate the crystallization onset in each white dwarf sequence.
\label{fig:Bsur}}
\end{figure*}

\subsection{Magnetic field evolution for different masses}

We computed the magnetic field evolution using the methodology described in Section \ref{sec:2.3} for the different WD models evolved with MESA. For each mass, we considered four cases of the magnetic field evolution to cover most of the physical processes expected during WD cooling and the expectations of the magnetic field geometry from observations. Thus, the cases covered two geometries of the field: dipole and dipole+quadrupole, each combined with two cases of diffusivity, only ohmic diffusion, and the other ohmic+turbulent diffusion. The latter case of diffusivity is used to consider turbulent diffusion driven by compositional convection during crystallization \citep{Castro-Tapia2024b} since many of the observed magnetic WDs with field strengths $\gtrsim10^{6}$ G have cooling ages larger that the time expected for the onset of crystallization \citep{2021MNRAS.507.5902B,2022ApJ...935L..12B}. 

In Figure \ref{fig:evolution} we show an example temporal evolution of the $B_{\mathrm{rms}}$ and diffusivity $\eta$ as a function of the radius for the $0.6\ M_{\odot}$ model assuming a dipole geometry and taking $\mathrm{f_{RM}}=10$ for turbulent diffusivity. This figure is similar to Figure 3 of \citet{Castro-Tapia2024b}. However, in the current case, we note how the magnetic field has already diffused to the surface of the WD when less than $5\%$ of the crystal core has formed as a direct consequence of the initial condition of the magnetic boundary computed by \citet{2024A&A...691L..21C} and letting the magnetic field evolve from the formation of the WD.

In order to later compare with observations, we computed the surface value of $B_{\mathrm{rms}}$ as a function of time from the magnetic field evolution of the models with different masses. In the left panel of Figure \ref{fig:Bsur}, we show $B_{\mathrm{rms,surf}}$ normalized to the initial value $B_0$, as a function of the cooling time when considering only ohmic diffusion for the different masses and d and d+q field geometries. We define the reduction factor $f_{B}(t_{\mathrm{cool}})=(B_{\mathrm{rms,sur}}/B_{0})(t_\mathrm{cool})$, which indicates the expected value of the magnetic field at the WD surface, normalized by the initial field, as a function of the WD cooling time. In the right panel, we show the same but including turbulent diffusion with $\mathrm{f_{RM}}=10$. Regardless of the mass, all the cases show that $B_{\mathrm{rms,surf}}$ increases as it is transported from the interior to the surface, reaching a maximum intensity, to then decay, as it is diffused out of the WD, with a much smaller rate of change than the first increase. This slow decay is a direct consequence of the crystallization process that increases the conductivities as the solid phase grows.

We note that for both panels of Figure \ref{fig:Bsur} the geometry d+q diffuses the magnetic field faster than the d geometry because of the more rapid decay of the quadrupole mode, which now forms part of the field structure \citep{Cumming2002}. The effect of turbulent diffusion shown in the right panel of the figure is to reduce the value $B_{\mathrm{rms,surf}}$ for about 1 Gyr after the crystallization onset, to change slowly again later. This is similar to what is observed in \citet{Castro-Tapia2024b}. However, in that work, $B_{\mathrm{rms,surf}}$ remains almost constant after reaching a maximum since the evolution of the magnetic field starts with the crystallization process; then, as the field is transported it becomes embedded as the solid phase reaches the outermost layers of the WD preventing the diffusion out of the star. We see something similar happening in the case of $0.6\ M_{\odot}$ dipole without turbulent diffusion, because its maximum $B_{\mathrm{rms,surf}}$ is reached after the onset of crystallization.

\section{Comparison with observations}

\begin{figure}
\includegraphics[width=0.99\columnwidth]{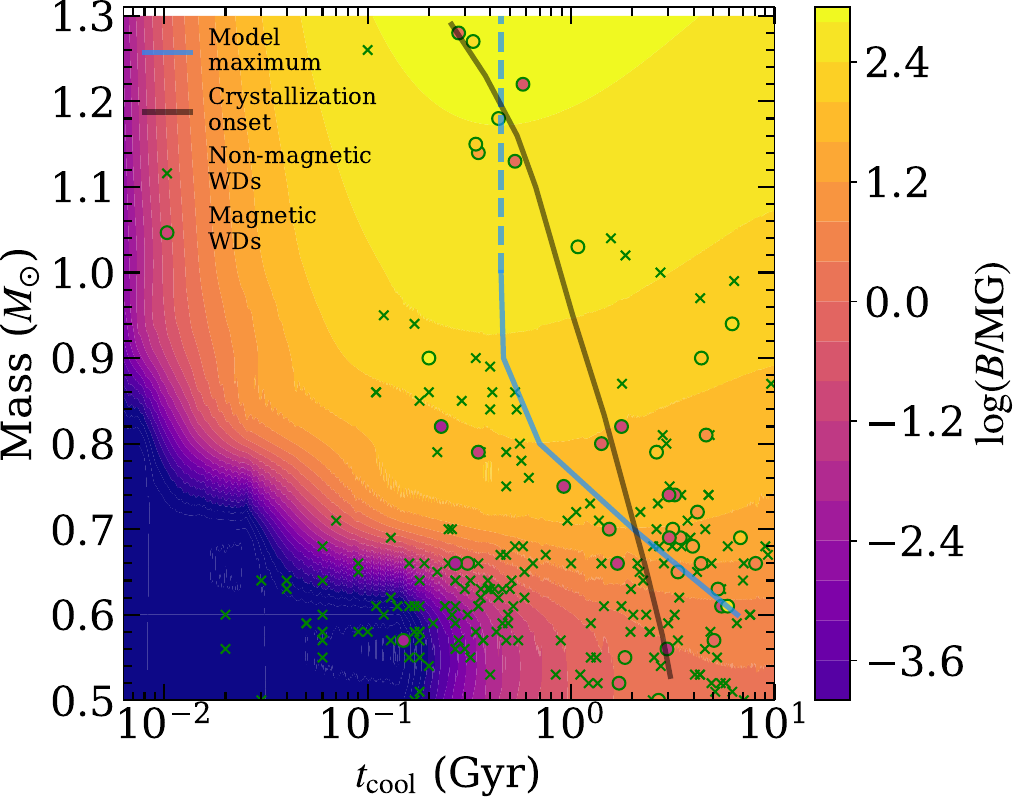}
\caption{Color map indicating the logarithmic prediction of the magnetic field at the WD surface as a function of the WD mass and cooling time. {\bf For this, we assumed dipole geometry, a $B_{0}$ scaled to equipartition from the main-sequence dynamo, and considered magnetic flux conservation}. For comparison, the non-MWDs and MWDs from the local 20-40pc sample of \cite{2022ApJ...935L..12B} are shown with the same color scale as the background for their measured field. The crystallization onset and the moment when the magnetic field models reach their maximum as a function of the WD mass are plotted using black and cyan lines, respectively (the dashed part of the cyan line takes the same age as the maximum of the $1\ M_{\odot}$ model for the more massive WDs).
\label{fig:emergence}}
\end{figure}

\begin{figure*}
\includegraphics[trim=0 0 0 0,clip,width=0.99\columnwidth]{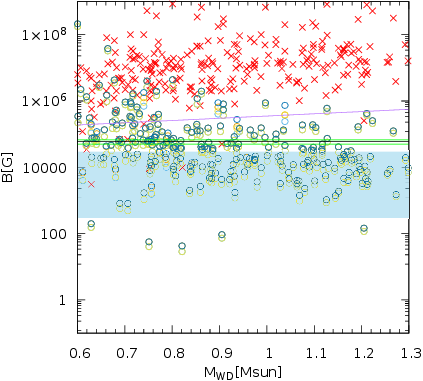}
\includegraphics[trim=0 0 0 0,clip,width=0.99\columnwidth]{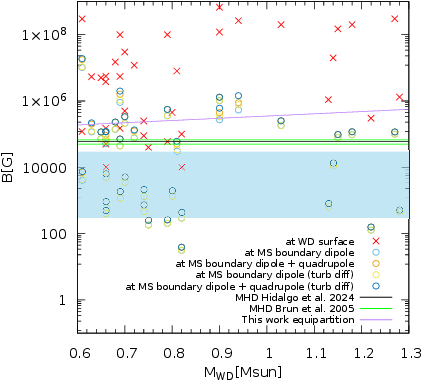}
\caption{Left panel: Magnetic fields at the surface of white dwarf stars taken from the Montreal White Dwarf Database (red crosses), and their corresponding magnetic fields expected at the edge of the magnetic boundary during the main sequence stage, assuming different considerations for the field geometry and diffusivities (circles). The solid lines indicate the magnetic fields expected at this magnetic boundary during the main sequence, taken from the MHD simulations of main sequence A stars of \cite{2005ApJ...629..461B} (green lines) and \cite{2024A&A...691A.326H} (black line) and from our estimations assuming equipartition between the kinetic and magnetic energy (purple line). The blue shaded area indicates the typical magnetic field strengths measured at the surface of magnetic main sequence Ap and Bp stars. Right panel: Same as left panel but for the sample of magnetic white dwarfs within 20pc from our Sun and magnetic white dwarfs younger than 0.6 Gyrs within 40pc from our Sun \citep{2022ApJ...935L..12B}.
\label{fig:observations}}
\end{figure*}

\begin{figure*}
\includegraphics[width=0.99\textwidth]{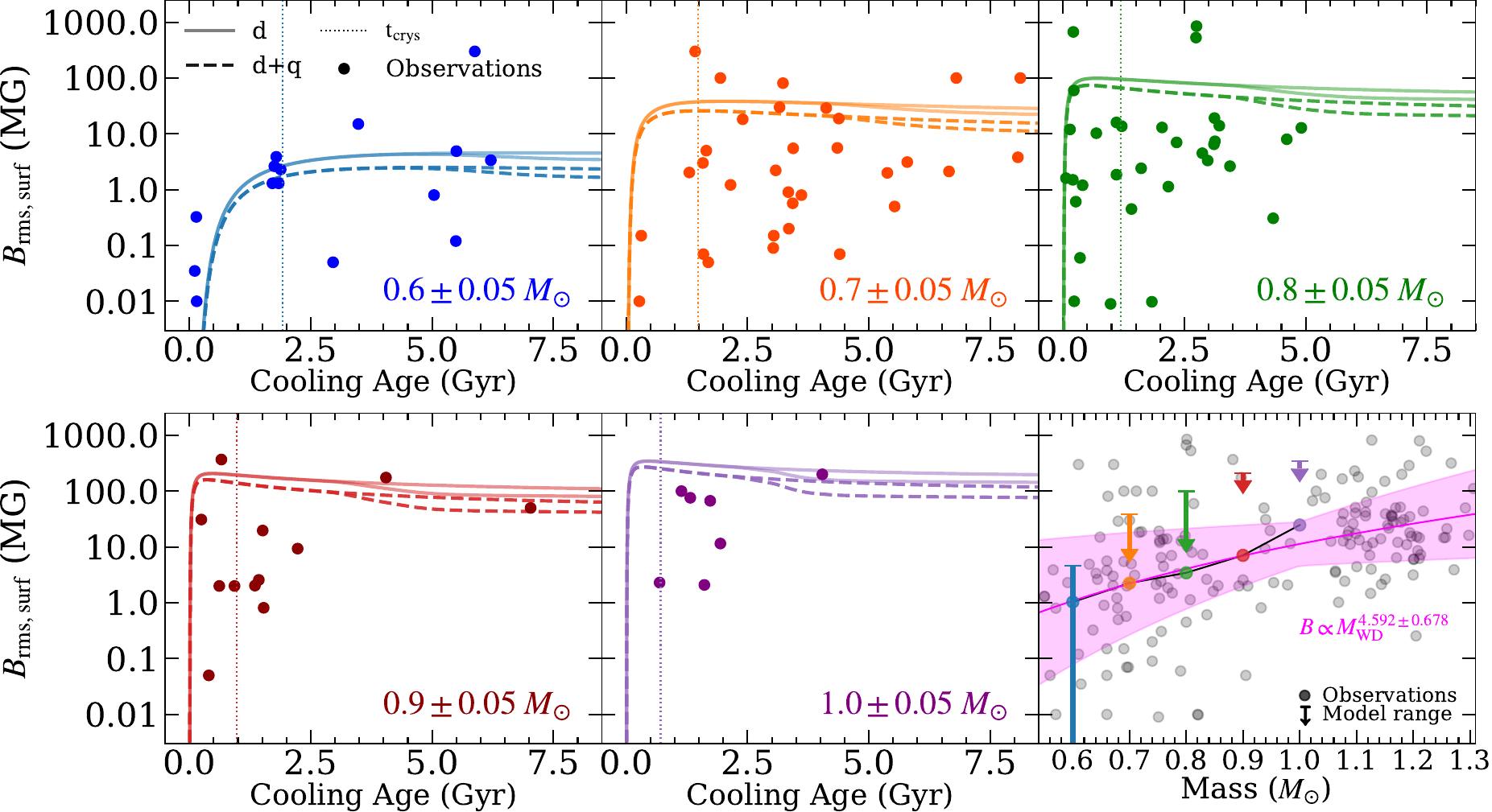}
\caption{Predictions of the surface magnetic field strengths (rms) as a function of cooling age for our models compared to observed magnetic fields in a 100 pc sample of the MWDD combined with the MWDs of \citet{2022ApJ...935L..12B}. The initial magnetic field prediction $B_{0}$ comes from equipartition in the main-sequence and rescaling assuming conservation of the magnetic flux (Eq. \ref{eq:MS_field}). We have separated into different panels the masses of the models in the range $0.6$-$1\ M_{\odot}$, and we compare in each panel with MWDs with masses within a range of $\pm0.05\ M_{\odot}$ for the corresponding model mass. The solid lines show the d geometry models, while the dashed lines are the d+q models. For each geometry, the split in the lines at some point of the evolution shows the effect of turbulent diffusion with $\mathrm{f_{R_{m}}}=10$, being the cases of more rapid decay the ones that include this factor. In the bottom-right panel, we summarize the range of the magnetic field strengths predicted and observed for different masses; for the observations used in the previous panels we show a logarithmic average of the magnetic fields in the bin, whereas for the model range, the upper limit is the maximum strength predicted, and the arrow is at the minimum $B_{\mathrm{rms, surf}}$ for the age of the youngest WD in the bin. A linear fit in the logarithmic space of mass and B (magenta line) for all the data points shown (in black) is included. The shaded (magenta) area shows the $\pm5\sigma$ range of the fit parameters fitted (see text).}
\label{fig:observations_2}
\end{figure*}

To compare our magnetic field evolution models with observed surface magnetic fields in WDs, we take the local 20-40 pc sample of WDs
from \citet{2022ApJ...935L..12B} and the MWDs available in the Montreal White Dwarf Database \citep[MWDD,][]{Dufour2017}. In Figure \ref{fig:emergence}, we show the mass and cooling age of the MWDs from the 20-40 pc sample. As a background color map, we show the logarithm of the $B_{\mathrm{rms,surf}}$ value for each mass and cooling age predicted by our dipole evolution models, where we assumed a $B_{0}$ for each mass scaled according to magnetic flux conservation as $B_{0}= B_{\mathrm{MS}}(R_{\mathrm{WD,b}}/R_{\mathrm{MS,b}})^{2}$, where $R_{\mathrm{MS,b}}$ and $R_{\mathrm{WD,b}}$ are the radius at the magnetic boundary of the main-sequence progenitor and the WD, respectively, and taking the equipartition prediction for $B_{\mathrm{MS}}$\footnote{As we only computed models for the magnetic field evolution of up to $1\ M_{\odot}$ WDs, for the colormap background of $>1\ M_{\odot}$  we used the same reduction factor predicted for the $1\ M_{\odot}$ case and extrapolated $B_{0}$ using a linear fit in the logarithmic space with the mass.}. The same color scaling is used to show the value of the observed surface magnetic field of the MWDs in the sample. This comparison confirms the results of \citet{2024A&A...691L..21C}: the late appearance of magnetic fields in WDs is not necessarily related to the crystallization process. For any fixed mass in Figure \ref{fig:emergence}, we see that the MWDs tend to accumulate around the point where the maximum of the model prediction is reached. We also see that most of the magnetic WDs are located in the regions where we expect the surface magnetic field to be larger. While the majority of MWDs appear near or after the onset of crystallization, those that appear before can only be explained by the appearance of a magnetic field formed before the WD phase, and may not be related to crystallization at all.

While the models shown in Figure \ref{fig:emergence} predict that for increasing mass, more MWDs must appear over a wide range of cooling age, we acknowledge that our models consider an ideal case, where the magnetic field generated during the main-sequence, with perfect equipartition assumption, reaches the WD phase without energy loss. Despite these caveats, we see that at about $>0.1$ Gyr, for the $\gtrsim1\ M_{\odot}$ models, the field predictions are of the same order of magnitude as the maximum magnetic field reached. Something similar happens for the less massive MWDs, but at $\gtrsim1$ Gyr. Moreover, following the color scaling, it seems that the observations are almost all at the same or below the value predicted by the background map, confirming that the models can be interpreted as the best-case scenario for the magnetic field formation and evolution.

In order to predict the value of the magnetic field in the main-sequence phase $B_{\mathrm{MS}}$ necessary to reproduce the observed surface fields in MWDs we consider the reduction factors $f_{B}(t_{\mathrm{cool}})$ from the different evolution models and the assumption of magnetic flux conservation given by $R_{\mathrm{MS,b}}$ and $R_{\mathrm{WD,b}}$. Thus, for a given observed field $B_{\mathrm{rms,surf}}$ in the WD phase, its corresponding magnetic field strength generated during the main-sequence can be computed as follows 
\begin{equation}\label{eq:MS_field}
B_{\mathrm{MS}}=\frac{B_{\mathrm{rms,obs}}}{f_{B}(t_{\mathrm{cool}})}\left(\frac{R_{\mathrm{MS,b}}}{R_{\mathrm{WD,b}}}\right)^{2},
\end{equation}
where $f_{B}(t_{\mathrm{cool}})$ is the prediction of our model at a given cooling time $t_\mathrm{cool}$ because in this case we do not need to know $B_{0}$ a priori.

In Figure \ref{fig:observations} we show the observed magnetic field in MWDs (red crosses) for different masses and their corresponding predictions of the main-sequence field $B_{\mathrm{MS}}$ (empty circles) necessary to reproduce the observations, obtained using equation \eqref{eq:MS_field} for our models with d and d+q geometries. We also included the prediction from our models with turbulent diffusion. In the left panel of the figure, we show the MWDs in the MWDD, while the right panel shows the MWDs in the 20-40 pc sample of \citet{2022ApJ...935L..12B}. To compare our $B_{\mathrm{MS}}$ predictions, we added the magnetic field strength prediction from a linear fit to the equipartition value at the magnetic boundary of main-sequence progenitors for different WDs, assuming an initial-to-final-mass relation from \citep{2008MNRAS.387.1693C}. We also compare the predicted magnetic field strengths from core convection in the main-sequence obtained from MHD simulations of \citet{2005ApJ...629..461B} and \citet{2024A&A...691A.326H}, for Main Sequence masses of 2 and 2.2M$_\odot$, respectively. 

We note that the necessary $B_{\mathrm{MS}}$ predictions to reproduce the observations are mostly in the range $\sim10^{3}$-$10^{5}$ G, which is very similar to the equipartition and MHD predictions of $\sim10^{4}$-$10^{5}$ G. The equipartition prediction appears almost as an upper limit for the main-sequence convective core dynamo field necessary to reproduce the observed MWDs with $\gtrsim0.75\ M_{\odot}$, except for a very few whose $B_{\mathrm{MS}}$ predictions are larger than the equipartition by a factor of $\lesssim 5$. For MWDs with $\lesssim 0.75\ M_{\odot}$, the $B_{\mathrm{MS}}$ predictions appear more scattered from the equipartition value, some of them even being the highest $B_{\mathrm{MS}}$ for all the masses. This occurs because of the small magnetic boundaries and larger radius for the less massive WDs, which lead to a maximum factor $f_{B}\lesssim 0.03$. Therefore, to reproduce MWDs with about the same few MG strength, the $B_{\mathrm{MS}}$ must be larger for the less massive stars.  However, for all the masses, about half of the magnetic fields observed in WDs could have originated from $B_{\mathrm{MS}}$ weaker than $10^{4}$ G, below the equipartition and MHD simulations prediction. Some MWDs with $10^{5}$-$10^{6}$ G could have even been produced with $B_{\mathrm{MS}}$ of just a hundred G. It is important to remark that the initial-to-final-mass relation and, consequently, the value of $R_{\mathrm{MS,b}}/R_{\mathrm{WD,b}}$ are subject to large uncertainties, particularly for the more massive WDs. Therefore, the value of $B_{\mathrm{MS}}$ could be much weaker or stronger than predicted here.

The inclusion of models with d and d+q geometry sets an error in the estimation of $B_{\mathrm{MS}}$. Depending on the cooling age of the WD, having a d+q geometry implies that the $B_{\mathrm{MS}}$ prediction is larger than the d case at most by a factor of $\sim2$. As mentioned in Section \ref{sec:3.1}, the d+q geometry can reproduce the off-centered dipole nature that fits many MWDs \citep{Hardy2023,Moss2025}, although the individual contributions of dipole and quadrupole components to the d+q model may depend on each MWD, and the initial conditions assumed here only set an approximate estimation for how this model is different from the simple dipole. While including a turbulent diffusion component only sets an error estimation for those MWDs that are already undergoing crystallization, for the oldest ones, the $B_{\mathrm{MS}}$ prediction can vary up to a factor of $\sim 4$ when contrasting the d+q geometry including turbulent diffusion with the d geometry without turbulent transport. 

For a comparison of the magnetic field strength observed in MWDs and the predicted from our models, we considered the magnetic field at the start of the WD phase $B_{0}$, to be the one predicted from magnetic flux conservation of the equipartition prediction of $B_{\mathrm{MS}}$, and considered the diffusion for about 9 Gyr. In Figure \ref{fig:observations_2}, we show the evolution of $B_{0}$ for the models with masses $0.6\ M_{\odot}$ to $1\ M_{\odot}$, considering the d and d+q geometries and turbulent diffusion and not. We compare these models with the observed fields of MWDs in a 100 pc sample with masses in the range $0.55$-$1.05\ M_{\odot}$ in bins of mass separated by $0.1\ M_{\odot}$, as shown for different panels of Figure \ref{fig:observations_2}. We note that the dipole model without turbulent diffusion appears almost as an upper limit for the observations for each mass. Additionally, we note that many of the observations seem to roughly follow the pattern given by the models: the magnetic field strength is increasing with time, almost saturating after the model reaches a maximum, again confirming that the late appearance of MWDs can still be explained by a main-sequence dynamo \citep{2024A&A...691L..21C}.

In principle, factors such as the rotation of the star should affect the scaling of the magnetic field when it is generated during the main-sequence. However, numerical simulations on convective dynamos show that the magnetic field can reach a significant fraction of equipartition \citep{2024A&A...691A.326H}, then it seems reasonable considering such a scaling as an upper limit. Other uncertainty sources in determining the magnetic boundary are related to the whole stellar evolution history \citep{2024A&A...691L..21C}, such as convective boundary determinations, mass loss rates during the giant phases, and the differences in the initial-to-final-mass relation when assuming different stellar evolution models \citep{2018ApJ...866...21C}. While we did not attempt to explore these uncertainties, it is worth mentioning that our theory is presented in an idealistic scenario, where the magnetic field survives almost intact inside the radiative zones of the star during the giant phases.

From Figure \ref{fig:Bsur} it can be inferred that, if $B_{\mathrm{0}}$ was almost the same for any WD mass, the observed magnetic fields would be weaker on average for less massive WDs due to what we predict from solving the evolution of the magnetic field. This behavior is even more prominent in the models in Figure \ref{fig:emergence} and \ref{fig:observations_2} because the equipartition prediction increases with the WD mass, and because the factor of radius reduction at the magnetic boundary, ${R_{\mathrm{MS,b}}}/{R_{\mathrm{WD,b}}}$, is larger for more massive stars. In the lower right panel of Figure \ref{fig:observations_2}, we compare the observed magnetic field strengths at each mass; for this, we show the data in the whole mass range of the combined samples of MWDs (black dots) and a data point for the logarithmic average of the magnetic fields in the bins used in the previous panels. On top of each bin, we plot the model range, where the upper limit is given by the maximum field strength predicted, and the arrow is at the minimum model value predicted for the age of the youngest WD in the bin. We see that the average magnetic field strengths in this 100 pc sample increase with mass, which is similar to what we predict with the evolution models.

As a more appropriate test of the apparent correlation between magnetic field strength and WD mass, we made a linear fit in the logarithmic space of mass and $B$ for all the data points shown (in black) of the combined samples, which means that we fit an exponential relation characterized by $B\propto M^{a\pm\sigma_{a}}_{\mathrm{WD}}$. We find a value $a=4.592\pm0.678$, showing a positive correlation between the mass and $B$, which is true for our fit even if we consider the case where the exponential factor is $a-5\sigma_{a}$. In the lower right panel of Figure \ref{fig:observations_2}, we also show the fit made (magenta line) considering the $\pm 5\sigma$ range of the fitted parameters (shaded magenta area). This result shows that despite the large dispersion in the data, the positive correlation between WD mass and magnetic field strength might be significant.

It has been reported in the literature that MWDs are, on average, more massive than non-MWDs \citep{2020AdSpR..66.1025F}. Whether more massive MWDs have stronger magnetic fields is still poorly constrained. Recently, \citet{Moss2025} identified 167 MWDs, with 87 new discoveries, within the 100 pc sample of \citet{Kilic2025}. In this new MWDs sample, it is reported that at least two groups are identified in a three-parameter space analysis, one group of older ($\sim2.9$ Gyr), less massive ($\sim0.7\ M_{\odot}$), and with weaker magnetic fields ($\sim 3.7$ MG); and other group of younger ($\sim1.8$ Gyr), more massive ($\sim0.96\ M_{\odot}$), and with stronger fields ($\sim 84$ MG). These results favor our analysis, since our models predict that stronger magnetic fields tend to appear earlier in more massive WDs.

Additionally, \citet{Moss2025} noted that more MWDs in their sample can be explained by the emergence of a main-sequence magnetic field during the WD phase than by the magnetic field generated through a crystallization-driven dynamo. However, they only used the breakout times estimated by \citet{2024A&A...691L..21C} and \citet{2024MNRAS.528.3153B}, and the strength of the magnetic field was not addressed to compare the theories. Here, we show how adding the magnetic field strength in the calculation may also favor the appearance of a main-sequence magnetic field during the WD phase to explain some magnetic MWDs of any mass. Conversely, \citet{Castro-Tapia2024b} predicted that the crystallization-driven dynamo theory can only explain MWDs of up to a few MG in the best-case scenario.

\section{Summary and Conclusions}

We have studied the evolution of WD magnetic fields that originated from core-convective dynamos during the main-sequence phase. For this, we used the estimation of \citet{2024A&A...691L..21C} for the magnetic boundary expected in the WD phase for different masses as an initial condition. It is important to recall that this magnetic boundary is subject to large uncertainties that arise both from the initial-to-final-mass relation and from the treatment of convective boundaries. We then take the methodology that was used in \citet{Castro-Tapia2024b} to evolve a uniform axisymmetric poloidal field. In this way, we were able to conduct a more insightful analysis of the surface magnetic field strength expected at different cooling ages and masses, assuming MWDs are consistent with a main-sequence core-convective dynamo origin. In general, we find that this theory can explain many observed MWDs for various ages, masses, and field strengths. We see that for all the WD models, the surface magnetic field strength increases, reaching a maximum, then decays slowly to remain almost constant after crystallization. Such behavior roughly resembles the observed WD fields.

We analyze two different geometries for the poloidal field: a simple dipole and a dipole plus a quadrupole component. We find that the latter is naturally consistent with an off-centered dipole model, which has been proposed to explain photometric and spectroscopic data of MWDs \citep{Hardy2023}. In doing so, we observe that the d+q geometry yields weaker predictions for the surface magnetic field strengths for a constant initial magnetic field $B_{0}$ in the WD phase. This occurs due to the more rapid decay of the quadrupole component in the magnetic field expansion \citep[e.g.][]{Cumming2002}. Additionally, we also consider models with turbulent diffusion during crystallization as in \citet{Castro-Tapia2024b}. Having these different models lead to surface magnetic field strengths that vary by a factor of 2 to 4 for the same cooling age and mass.

To assess the feasibility of the main sequence core-convective dynamo in terms of magnetic field strength that would be observed in the WD phase, we take the strength observed in two samples of MWDs and trace back the necessary strength produced during the main-sequence to reproduce the observed fields. For this, we also consider conservation of the magnetic flux for the contraction of the magnetic boundary from the main-sequence to the WD phase. We compare the back-tracked main-sequence magnetic field with MHD simulations of the convective-core dynamo and with the equipartition prediction from the core convective velocities, and find that most of the back-tracked fields appear scattered around or below the values of the MHD and equipartition predictions for masses $\gtrsim0.75\ M_{\odot}$.

For WDs with $\lesssim0.75\ M_{\odot}$, we find more back-tracked fields appear above the MHD and equipartition prediction, which is a direct consequence of the smaller magnetic boundary predicted by \citet{2024A&A...691L..21C} for the less massive WDs. As more diffusion must occur to reach the surface for deeper magnetic boundaries, this results in a smaller portion of the magnetic field reaching the surface before the inner parts become frozen-in by the larger conductivities reached during crystallization \citep{Castro-Tapia2024b}. Additionally, it is not clear if the more extended convective zones for less massive stars during central He burning would vanish the field generated during the main-sequence or power it. Therefore, according to \citet{2024A&A...691L..21C}, there is more uncertainty for the feasibility of the main-sequence core-convective dynamo mechanism for the origin of MWDs with $\lesssim0.65\ M_{\odot}$.

By taking a sample limited to 100 pc in the Montreal WD Database, combined with the MWDs in \citet{2022ApJ...935L..12B}, in the range of 0.55 to 1.05 $M_{\odot}$ and the prediction of $B_{0}$ from equipartition for our evolution models, we compare the field strength as a function of time for fixed masses. We find that the models with different geometries and with and without turbulent diffusion cover a range that can fit many of the strengths observed in MWDs. However, in general, using the equipartition assumption for $B_{0}$ appears almost as an upper limit for the field prediction at different masses, which would be expected if the magnetic field given by the core-convective dynamo is partially diffused during stellar evolution. \citet{MohapatraBlackman2024} recently pointed out that the core-convective dynamo may be unlikely if turbulent diffusion acts to rapidly diffuse the magnetic field when a stage of no magnetic field generation occurs during convection. They estimate the time for turbulent diffusion to be only $\gtrsim10$ yr, but only assumed that the field would dissipate in the convection zone. Conversely, \citet{2024A&A...691A.326H} showed that only a small fraction of the magnetic energy can reach the surface as diffusion acts in the region outside the convective core (even with similar turbulent diffusion times); then, more detailed MHD simulations of convection in the different stellar evolution stages are necessary to understand the survival and evolution of the core-convective generated magnetic field.

In light of the current observational features of MWDs, some studies point towards a multiple-channel magnetic field generation depending on mass \citep{2021MNRAS.507.5902B,2022ApJ...935L..12B,Moss2025}. This is because the late appearance of magnetic fields seems to be in line with the crystallization process for less massive WDs \citep{2024MNRAS.528.3153B} but not following a clear trend for ultramassive WDs \citep{2024MNRAS.533L..13B}. However, in this work, we show that observed magnetic field strengths seem to be roughly increasing with mass, which aligns with the predictions in our magnetic field evolution of core-convective dynamo fields that survived up to the WD phase. Additionally, in a recent study by \citet{Moss2025}, a similar feature is observed for a new sample of MWDs: the more magnetic WDs seem to accumulate towards higher masses.

We also follow up on the prediction of \citet{2024A&A...691L..21C} and show that, with a more detailed evolution of the magnetic field, the late appearance of high magnetic field strengths is not necessarily related to crystallization and may be explained almost solely by the appearance of a deep-seated magnetic field that was generated during the main-sequence and survived until the WD phase. Furthermore, such a theory may also explain MWDs with masses $\lesssim1.1\ M_{\odot}$ that appear before the crystallization onset, which, in the limited 100 pc that we use, tend to be weaker than those that appear after crystallization. However, note that this behavior is better explained if we take the cooling age at which our models reach a maximum surface magnetic strength value for each mass (Figure \ref{fig:observations_2}). The observed magnetic fields that are younger than the age at which the maximum model magnetic field is reached tend to be weaker than those that are older than that point.

In summary, while the survival of a magnetic field generated by a core-convective dynamo during the main-sequence is still under debate, if this field does survive, it could explain different features of observed magnetic fields in current WD samples; that is, more massive MWDs are younger and with higher magnetic field strengths than less massive MWDs \citep{Moss2025}. Many other issues must be addressed to fully understand the MWDs puzzle. First, the boundary where the magnetic field is expected to be trapped is not clear, as it depend on two of the major uncertainties in stellar evolution: the treatment of convective mixing and the initial-to-final-mass relation. Second, in our models, we employ different geometries and diffusion treatments to account for errors in the estimation of the evolved magnetic field when comparing with various samples. Nevertheless, a case-by-case study is necessary to establish the correct setup for better constraining the origin of each MWD. 

Another source of uncertainty is the true nature of the core composition of WDs with masses $>1.1\ M_{\odot}$. We have extrapolated the evolution of a $1\ M_{\odot}$ WD model with a C/O core to estimate the back-tracked magnetic field for ultramassive WDs in this work. By itself, this would not provide a very different result if we had taken the magnetic boundary for WDs with masses $>1.1\ M_{\odot}$ from \citet{2024A&A...691L..21C}, since the magnetic boundary remains very stable with values $M_{\mathrm{MB}}/M_{\mathrm{WD}}\gtrsim0.95$ for these masses, even when taking C/O or O/Ne cores. However, for ultramassive WDs that are likely to be formed during a post-main-sequence merger process \citep[e.g.][]{2012ApJ...749...25G} it is unknown what could happen to a magnetic field formed during the main-sequence. A recent work by \citet{BlatmanRui2025} has shown the potential of using pulsations and superficial magnetic fields to constrain the core composition of ultramassive WDs, although they only account for C/O and O/Ne cores with chemical profiles from single stellar evolution and did not compare with merger products, whose pulsating properties are different \citep{Althaus2021}. By studying magnetic fields in WDs, it is possible to indirectly study magnetic fields in earlier evolutionary stages, which will eventually help establish constraints on merger episodes. 


\begin{acknowledgments}
M.C.-T. thanks J.R. Fuentes, for his hospitality during a visit to CU Boulder. M.C.-T. is supported by the Fonds de recherche du Québec - Nature et technologies through a doctoral scholarship (\href{	https://doi.org/10.69777/366094}{DOI:10.69777/366094}).
M. C. acknowledges grant RYC2021-032721-I, funded by MCIN/AEI/10.13039/501100011033 and by the European
Union NextGenerationEU/PRTR.
This work was partially supported by the MINECO grant  PID2023-148661NB-I00 and by the AGAUR/Generalitat de Catalunya grant SGR-386/2021. M.C.-T. and S. Z. are members of the Centre de Recherche en Astrophysique du Québec (CRAQ). 
\end{acknowledgments}

%

\vspace{5mm}






\bibliographystyle{aasjournal}



\end{document}